\documentclass{webofc}
\usepackage{listings}
\usepackage[draft]{minted}
\usepackage{hyperref}
\usepackage{graphicx}
\usepackage{subcaption}
\usepackage[utf8]{inputenc}
\usepackage{tikz}

\usepackage[varg]{txfonts}

\newminted{cpp}{gobble=2, numbersep=3pt, escapeinside=||, xleftmargin=5mm, framesep=2mm, frame=leftline, baselinestretch=1, fontsize=\scriptsize, linenos=true}
\newmintinline{cpp}{}

\makeatletter
\AtBeginEnvironment{minted}{\dontdofcolorbox}
\def\dontdofcolorbox{\renewcommand\fcolorbox[4][]{##4}}
\makeatother

\usetikzlibrary{shadows, arrows, fit, positioning, matrix, shapes}
\usetikzlibrary{decorations.markings}

\definecolor{my_yellow}{RGB}{255, 253, 217}
\definecolor{my_orange}{RGB}{255, 127, 0}
\definecolor{my_lightblue}{RGB}{105, 186, 249}
\definecolor{my_purple}{RGB}{150, 154, 219}
\definecolor{my_green}{RGB}{90, 194, 160}

\tikzset {
  bigbox/.style = {draw, thick, fill=gray!10, rounded corners, rectangle},
  box/.style = {draw, thick, minimum height=0.8cm, minimum width=1.5cm, rounded corners, rectangle, fill=white, anchor=south},
  model/.style = {draw, thick, fill=white, text centered, minimum height=3em, minimum width=4em, rounded corners, drop shadow},
  user/.style = {draw, thick, ellipse, fill=white, text centered, minimum height=3em, minimum width=5em, drop shadow},
  line/.style = {->, thick, color=black, shorten <=2pt, shorten >=2pt, >=stealth'},
  dashedline/.style = {->, thick, dashed, color=black, shorten <=2pt, shorten >=2pt, >=stealth'},
  plain/.style = {minimum width=1em},
  arcnode/.style 2 args={
    decoration={
                 raise=#1,             
                 markings,   
                 mark=at position 0.5 with {\node[inner sep=0] {#2};}
            },
            postaction={decorate}
    }
}

\usetikzlibrary{shadows, calc}

\def\shadowshift{3pt,-3pt}
\def\shadowradius{6pt}

\colorlet{innercolor}{black!30}
\colorlet{outercolor}{gray!05}

\newcommand\drawshadow[1]{
  \begin{pgfonlayer}{shadow}
    \shade[outercolor,inner color=innercolor,outer color=outercolor] ($(#1.south west)+(\shadowshift)+(\shadowradius/2,\shadowradius/2)$) circle (\shadowradius);
    \shade[outercolor,inner color=innercolor,outer color=outercolor] ($(#1.north west)+(\shadowshift)+(\shadowradius/2,-\shadowradius/2)$) circle (\shadowradius);
    \shade[outercolor,inner color=innercolor,outer color=outercolor] ($(#1.south east)+(\shadowshift)+(-\shadowradius/2,\shadowradius/2)$) circle (\shadowradius);
    \shade[outercolor,inner color=innercolor,outer color=outercolor] ($(#1.north east)+(\shadowshift)+(-\shadowradius/2,-\shadowradius/2)$) circle (\shadowradius);
    \shade[top color=innercolor,bottom color=outercolor] ($(#1.south west)+(\shadowshift)+(\shadowradius/2,-\shadowradius/2)$) rectangle ($(#1.south east)+(\shadowshift)+(-\shadowradius/2,\shadowradius/2)$);
    \shade[left color=innercolor,right color=outercolor] ($(#1.south east)+(\shadowshift)+(-\shadowradius/2,\shadowradius/2)$) rectangle ($(#1.north east)+(\shadowshift)+(\shadowradius/2,-\shadowradius/2)$);
    \shade[bottom color=innercolor,top color=outercolor] ($(#1.north west)+(\shadowshift)+(\shadowradius/2,-\shadowradius/2)$) rectangle ($(#1.north east)+(\shadowshift)+(-\shadowradius/2,\shadowradius/2)$);
    \shade[outercolor,right color=innercolor,left color=outercolor] ($(#1.south west)+(\shadowshift)+(-\shadowradius/2,\shadowradius/2)$) rectangle ($(#1.north west)+(\shadowshift)+(\shadowradius/2,-\shadowradius/2)$);
    \filldraw ($(#1.south west)+(\shadowshift)+(\shadowradius/2,\shadowradius/2)$) rectangle ($(#1.north east)+(\shadowshift)-(\shadowradius/2,\shadowradius/2)$);
  \end{pgfonlayer}
}

\newsavebox\mybox
\newlength\mylen

\newcommand\shadowimage[2][]{%
\setbox0=\hbox{\includegraphics[#1]{#2}}
\setlength\mylen{\wd0}
\ifnum\mylen<\ht0
\setlength\mylen{\ht0}
\fi
\divide \mylen by 120
\def\shadowshift{\mylen,-\mylen}
\def\shadowradius{\the\dimexpr\mylen+\mylen+\mylen\relax}
\begin{tikzpicture}
  \node[fill=white, rectangle, rounded corners, anchor=south west, inner sep=0] (image) at (0,0) {\includegraphics[#1]{#2}};
  \drawshadow{image}
\end{tikzpicture}}

\pgfdeclarelayer{background}
\pgfdeclarelayer{foreground}
\pgfdeclarelayer{shadow}
\pgfsetlayers{shadow,background,main,foreground}

\begin{document}

\title{C++ Modules in ROOT and Beyond}
        \author{\firstname{Vassil}
        \lastname{Vassilev}\inst{1}\fnsep\thanks
        {\email{vvasilev@cern.ch}}
        \firstname{David}
        \lastname{Lange}\inst{1}\fnsep\thanks
        {\email{david.lange@cern.ch}}
        \firstname{Malik Shahzad}
        \lastname{Muzaffar}\inst{2}\fnsep\thanks
        {\email{shahzad.malik.muzaffar@cern.ch}}
        \firstname{Mircho}
        \lastname{Rodozov}\inst{5}\fnsep\thanks
        {\email{mircho.nikolaev.rodozov@cern.ch}}
        \firstname{\nobreak{Oksana}}
        \lastname{Shadura}\inst{3}\fnsep\thanks
        {\email{oksana.shadura@cern.ch}} \and
        \firstname{Alexander}
        \lastname{Penev}\inst{4}\fnsep\thanks
        {\email{apenev@uni-plovdiv.bg}}
}

\institute{Princeton University, Princeton, New Jersey 08544, United States
\and
           CERN, Meyrin 1211, Geneve, Switzerland
\and
           University of Nebraska Lincoln, 1400 R St, Lincoln, NE 68588, United States
\and
           University of Plovdiv, Paisii Hilendarski, 236 Bulgaria Blvd, 4000 Plovdiv, Bulgaria
\and       Bulgarian Academy of Sciences, Sofia, Bulgaria
}

\abstract{
C++ Modules, one of the new features of C++20, aim to fix the long-standing build scalability problems in the language. They provide an IO-efficient, on-disk representation capable to reduce build times and peak memory usage. ROOT already employs the C++ modules technology in its dictionary system to improve performance and reduce the memory footprint.

ROOT with C++ Modules was released as a technology preview in fall 2018, after intensive development during the previous few years. The current state is ready for production, however, there is still room for performance optimizations. In this talk, we show the road map for making this technology enabled by default in ROOT. We demonstrate a global module indexing optimization which allows reducing the memory footprint dramatically for many workflows. We will report user feedback on the migration to ROOT with C++ Modules.
}
\maketitle
\section{Introduction} \label{intro}

 In C++ projects, the cost of header re-parsing is typically negligible in small to medium size codebases, but can be critical in larger codebases. Usually, scalability issues arise at compile-time and do not affect the programs at runtime. However, ROOT-based applications are different, as the ROOT C++ interpreter, Cling~\cite{Vasilev2012Cling}, processes code at program execution time. Therefore, avoiding redundant content re-parsing can yield better runtime performance~\cite{takahashi2019migratingToModules}. C++ Modules~\cite{vassilev2017Modules} have been supported in ROOT since version 6.16, in order to avoid unnecessary re-parsing and to improve the performance of applications using ROOT.

In~\cite{takahashi2019migratingToModules} we described the code sanitizing and infrastructure scaffolding required to fully employ the C++ modules technology on a large codebase. We outlined the set of challenges to be overcome in order to "modularize" the production code base of the CMS experiment, CMSSW, and showed preliminary performance results. That work was based on a beta version of the C++ modules technology in ROOT version 6.16.

In version 6.18 ROOT the technology matured and became default for version 6.20 for the Unix platforms. In this paper we describe recent performance results and suggest a more efficient strategy of querying information from C++ modules which yields better memory usage and replaces the eager loading of all modules at program initialization time.

\section{Background}

C++ Modules represent the compiler internal state serialized on disk, and can be deserialized on demand to avoid repetitions of parsing operations. This is described in detail in~\cite{vassilev2017Modules}. There are different implementations of the concept in Clang, GCC~\cite{GccCxxModules}, and MSVC~\cite{MsvcCxxModules}.

ROOT~\cite{Brun1997ROOT} has several features that are intended to allow rapid application development. They include automatic discovery of C++ headers that must be included, and the automatic loading required shared libraries. These features require a lot of computational resources by design. For instance, the underlying infrastructure must know the content of every reachable header file and library to properly function. ROOT uses Clang, through its interpreter Cling, and adopts a C++ Modules implementation to optimize out the need of unnecessary header file re-parsing~\cite{ClangModulesWeb}. 

Over the past several years ROOT has grown several levels of custom optimizations to prevent eager loading of header files and libraries which is vital for reducing the overall memory footprint. It is also important for improving execution speed, as Cling also processes code at runtime. Custom optimizations can be divided into three distinct types:
\begin{itemize}
    \item Reduction of feature usage -- performance-critical code should minimize its dependence on the automatic discovery feature. This is a very efficient optimization, however it relies on developers to carefully select which ROOT interface to use. Usually this is challenging for the developer due to the rich API that ROOT provides. In addition, relying on API internal implementations is fragile because it can silently break in future with changes in ROOT or other third-party code that the developer's codebase or ROOT depends on.
    \item Delay usage until needed -- ROOT implements "ROOTMAP" files containing a constructs to load the corresponding C++ header file when a definition from it is required.
    \item Precompute meta-information at compile time -- information about I/O streaming functions can be computed at dictionary compilation time and stored into files (denoted as "RDICT" files~\cite{OverviewModulesInROOT}). This captures ROOT's meta layer's state and makes runtime parsing of headers for I/O redundant in many cases. In addition ROOT caches heavily used header files in a precompiled header files (PCH) using in a highly efficient compiler optimization data structure.
\end{itemize}

There are several drawbacks of the existing infrastructure. Firstly, the maintenance is done by a small community. Secondly, the PCH optimization can not be applied to third-party codebases, would be beneficial for many cases. Thirdly, the implementation of ROOTMAP-based~\cite{OverviewModulesInROOT} infrastructure has many deficiencies and its correctness is questionable due to the many existing bugs.

The C++ modules implementation in ROOT targets minimizing maintenance costs by relying on features being developed and maintained by the Clang community. It also provides compiler-grade correctness and PCH efficiency. ROOT builds a module file (PCM) per library as opposed to a PCH per process. Loading PCM files is very efficient but introduces a constant memory overhead depending on the size and complexity of the module content. In an example application, the loading of all module files at initialization time increases the memory overhead for ROOT standalone by 29\% and embedded in CMSSW with 2800\% in ROOT version 6.20. Although, in the long term the preloading of module files is likely to become more efficient and converge towards being a zero-overhead operation, we can provide a more efficient approach with the infrastructure in place already today.

%
%
%

\section{Indexing Module File Contents}

The Clang compiler (thus Cling) queries information from C++ module files by extending the name lookup. Every time a new identifier is processed, the compiler looks it up by name to check if it exists and to tabulate its properties. In case the identifier is known to a module its containing entity is streamed from disk into memory. If the identifier is known by more than one module, the information from all relevant modules is deserialized into memory and the compiler does further work to de-duplicate possible content overlap.

In practice, registering the set of identifiers exported by each module file is sufficient to model the C++ language semantics. However, the implementation does a few more operations such as pre-allocating source locations for diagnostics, deserializing class template specializations and emitting virtual tables for classes with inlined key functions. These operations are implementation deficiencies and are very likely to be resolved in future versions. Unfortunately, these deficiencies introduce a linear overhead ($O(N)$ where N is the number of modules) when all modules are loaded at initialization time. The PCH shares the same implementation but its overhead is $O(1)$. One possible way to address this implementation deficiency is to build a map of identifiers to their containing module.

The global module index (GMI) is an efficient on-disk mapping between identifiers and modules. It contains information about a set of modules that contain a given identifier. This data structure can be loaded at initialization time and used as a utility to efficiency load C++ modules on demand. There is an available implementation of the SGMI in Clang, which allows ROOT to reduce its initialization time and lazily load the necessary C++ modules. 

\begin{figure}[H]
\centering
\begin{subfigure}[t]{.49\linewidth}
\shadowimage[width=.9\textwidth]{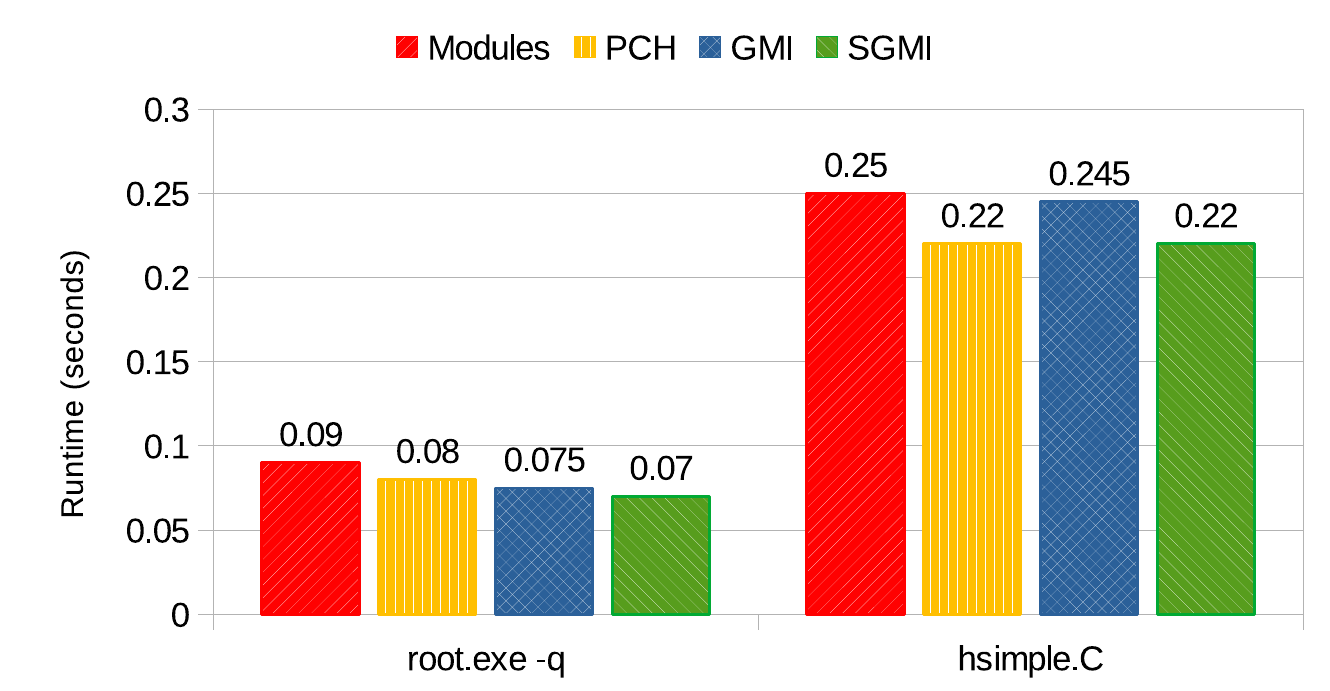}
\caption{Runtime of workflows short execution time} \label{fig:perf:1a}
\end{subfigure}
\begin{subfigure}[t]{.49\linewidth}
\shadowimage[width=.9\textwidth]{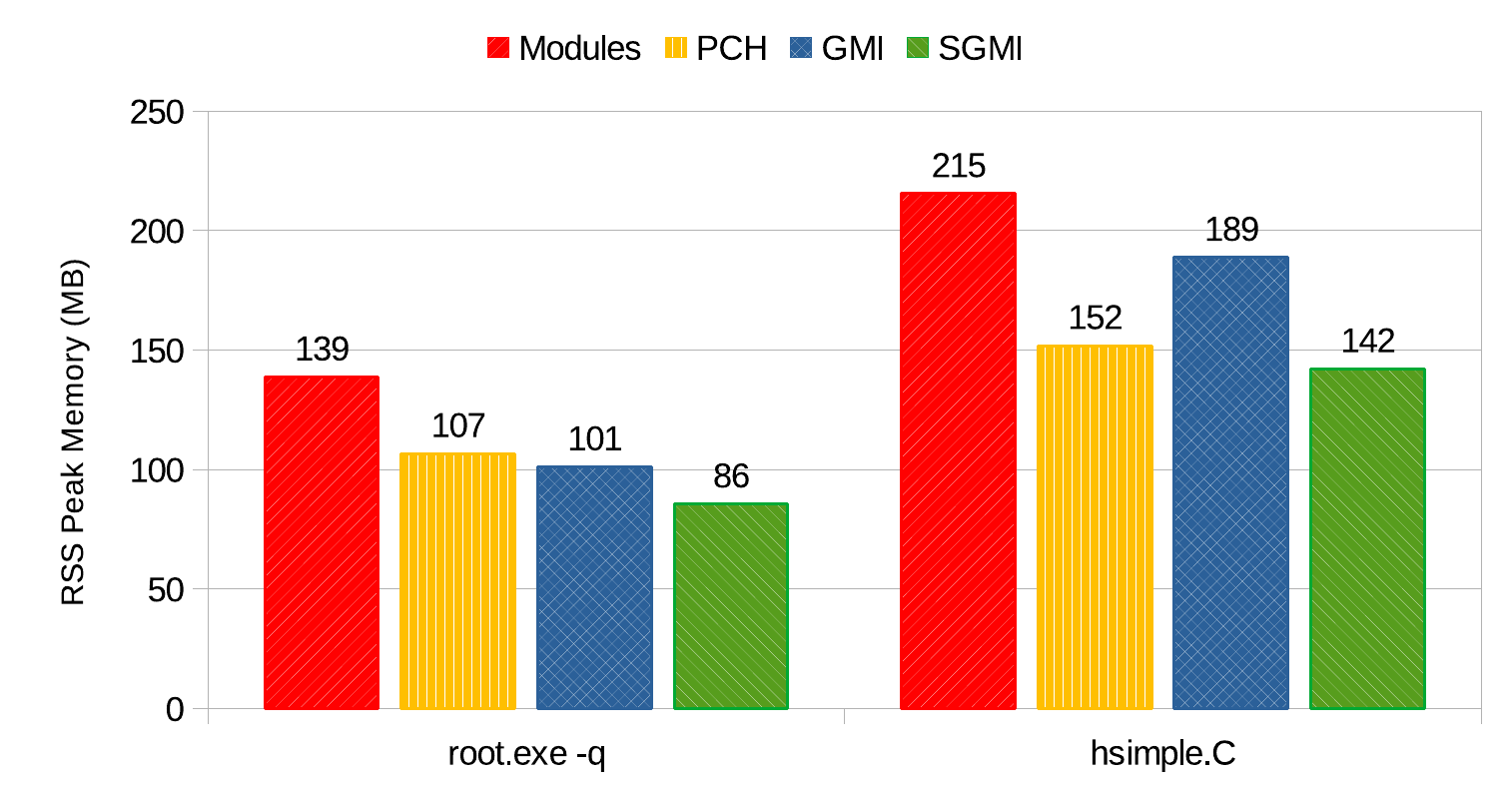}
\caption{Memory footprint of workflows with short execution time} \label{fig:perf:1b}
\end{subfigure}
\begin{subfigure}[t]{.49\linewidth}
\shadowimage[width=.9\textwidth]{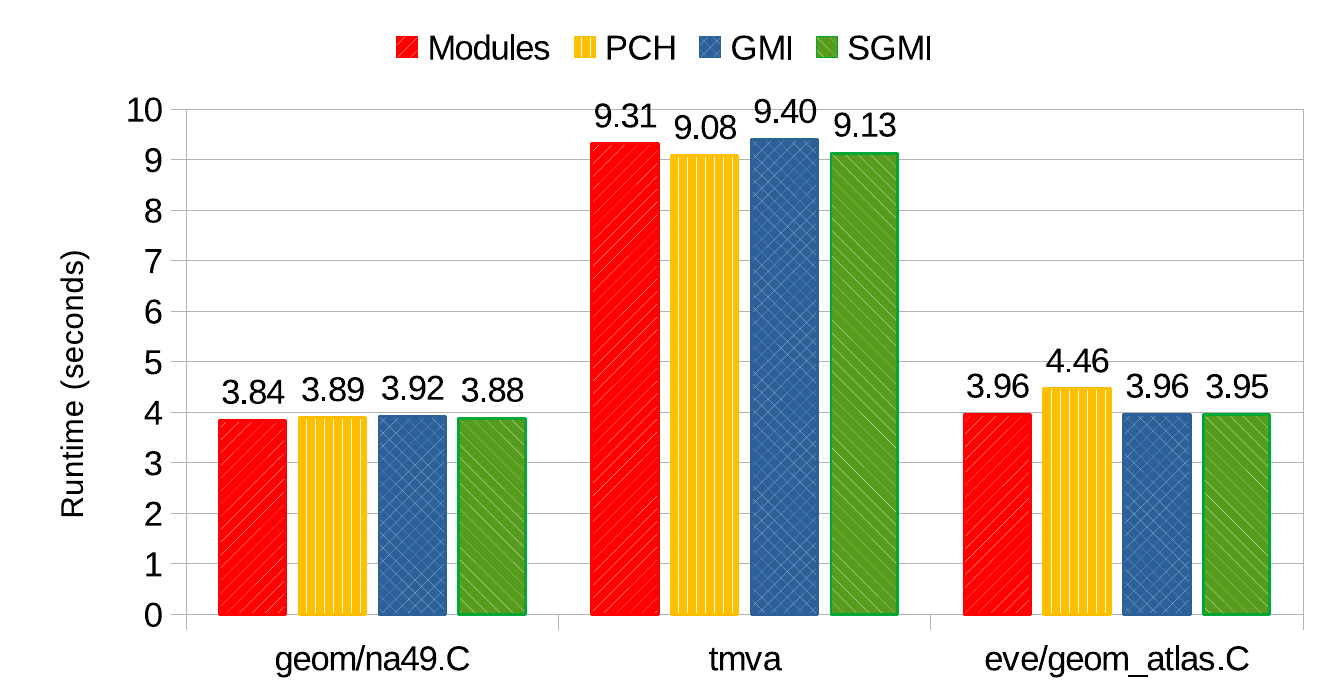}
\caption{Runtime of longer workflows} \label{fig:perf:2a}
\end{subfigure}
\begin{subfigure}[t]{.49\linewidth}
\shadowimage[width=.9\textwidth]{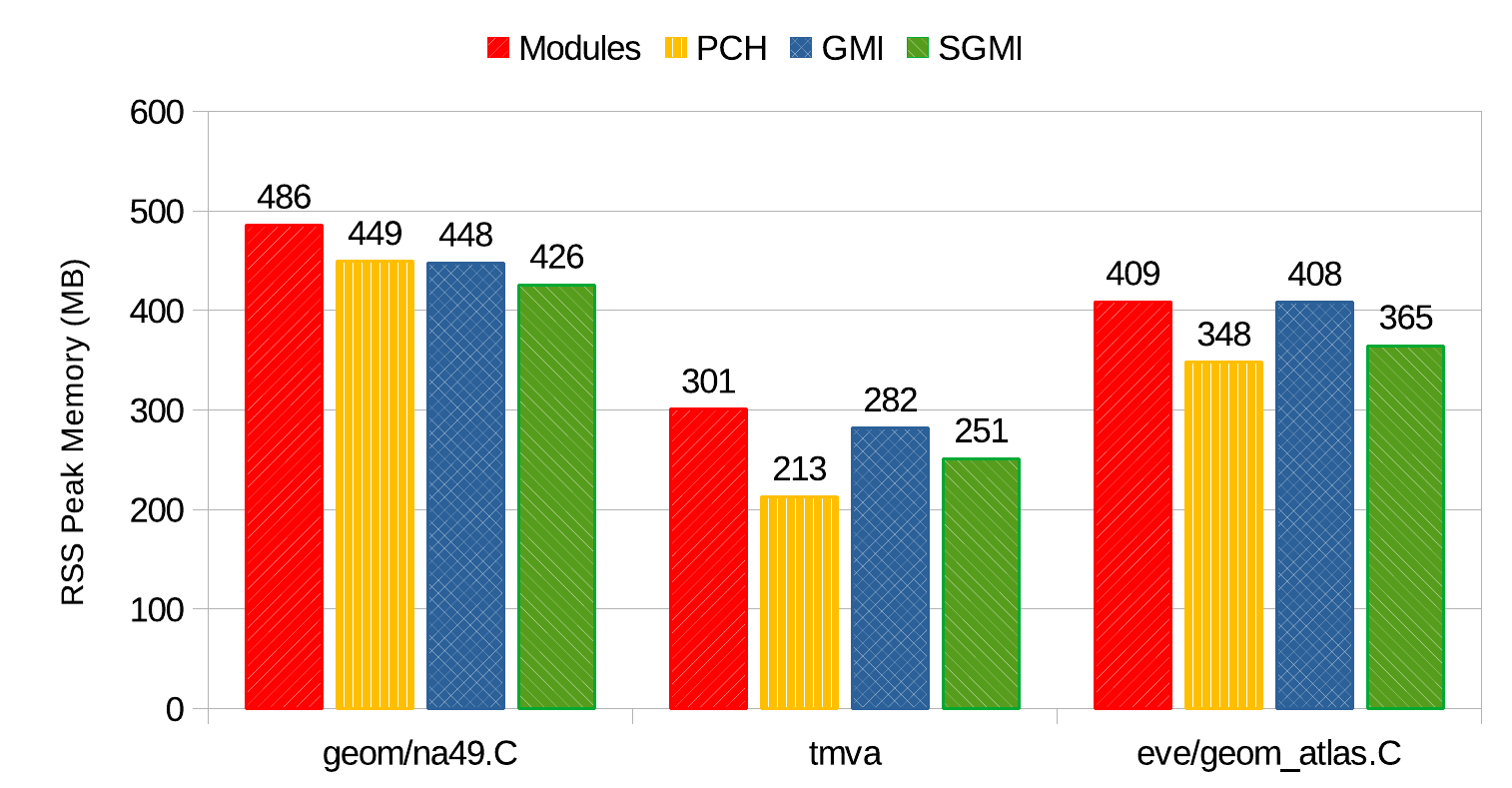}
\caption{Memory footprint of longer workflows} \label{fig:perf:2b}
\end{subfigure}
\caption{Performance benchmarks in ROOT}\label{fig:gmiperf}
\end{figure}

The GMI is purely lexical. For example upon information query of \emph{Gpad}, the GMI will return several modules that contain that identifier. Only one module contains the \emph{Gpad} definition and the rest contain forward declarations which are not used but result in loading of the corresponding C++ module file. A relatively cheap implementation improvement is to return only the module containing a defining entity given the identifier. A semantic GMI (SGMI) implementation is capable of reducing the false positive C++ module loads. The implementation keeps track of the module which defines the entity referred by an identifier. 

Figure~\ref{fig:gmiperf} shows the performance comparisons of the loading all C++ Modules at initialization time (red), using a PCH (yellow), lexical global module index (blue), and the non-lexical new global module index (green). Lower is better.

We measured two sets of workflows: shorter workflows which work in millisecond time spans, and larger workflows working on a seconds time span. The shorter workflows intend to outline the cost of initialization time of ROOT using the different technologies and the larger workflows give some hints how the technology scales. The left-hand plot of these figures (Figure~\ref{fig:perf:1a} and~\ref{fig:perf:2a}) shows the runtime performance results. The right-hand side shows the memory footprint comparisons. The results are produced by the \emph{time -v} command which shows the kernel understanding of the CPU and memory usage. The plots in Figure~\ref{fig:gmiperf} were performed using Ubuntu 18.04, kernel 4.15.0-88-generic, i7-8809G Processor, 2x16 GB DDR4 2666, 1xSSD 512 GB (Intel NUC Hades Canyon 2018).

The \emph{root.exe -q -l -b} workflow starts up and immediately shuts down ROOT. The \emph{hsimple.C} workflow runs a minimal I/O example in ROOT. The \emph{geom/na49.C} runs a geometry tutorial which is sparsely optimized by the PCH. The \emph{tmva} runs a CPU-intense tutorial. The \emph{eve/geom\_atlas.C} is a gui-based sparsely-PCH-optimized tutorial.

Figure~\ref{fig:gmiperf} demonstrates that loading of all C++ module files at initialization time contributes to the overall peak memory usage, compared to the PCHs while it does not show significant run-time improvements. The expected suboptimal behavior can be mitigated by loading module files on demand. In number of cases the lexical GMI reaches the performance of the PCH. A smarter GMI implementation reaches the PCH level in most of the cases and in some cases outperforms it.

We take the memory results with a pinch of salt because they are not yet reconfirmed by regular memory analyzers such as \emph{heaptrack} and the allocation statistics in Clang. For example, \emph{heaptrack}, shows that the memory footprint of the semantic GMI is significantly better and the C++ modules results are better in general. As of writing of this paper the authors' understanding is that the heap allocations are less in size but greater in frequency which triggers some effect in the underlying allocator or low-level kernel allocation primitives to pre-allocate bigger chunks of memory.

\section{Integration in the CMS Software -- CMSSW}\label{cmssw}

The software for the high-level trigger and offline reconstruction of the CMS experiment, \emph{CMSSW}, poses two different challenges with respect to C++ modules in ROOT: C++ module file relocation and incremental development. CMSSW deploys ROOT in computing centers which requires the C++ module files to be relocatable. In addition, the releases of CMSSW are typically distributed in a non-writable network file system \emph{cvmfs}\cite{cvmfs,muzaffar2010optimization}. Programmers clone a package locally to make modifications of CMSSW algorithms. In turn, the locally cloned package takes precedence over the one in the release area. The relatively straight-forward design makes it challenging for the C++ modules infrastructure to produce compatible files because of the possible module file cross references. In addition, the on-demand creation of GMI is problematic due to its requirement to be in the same folder as the base release (a limitation as currently implemented). A solution to this problem is to generate the GMI at build time and then implement some logic to exclude the locally cloned packages from module file resolution.


As is the case when building ROOT, CMSSW module definitions are stored in text files where one module definition corresponds to one library. Module definitions are generated by the SCRAM \cite{SCRAM,muzaffar2010optimization} build system as part of the build process. Each library stores the module definition description into \emph{Library\_Name.modulemap}, then all \emph{modulemap} files are concatenated into a final \emph{module.modulemap} file. Initial work has focused on generating module definitions for the libraries which require ROOT I/O information (dictionaries). However, we have determined that that a number of problems that we have encountered can be fixed by generating module definitions for transitively included header files from the dictionary. For example, if the dictionary file includes header files from \emph{CLHEP} or \emph{tinyxml} we should have a module for them as well. This will also avoid problems, either compilation errors or loss of performance, from duplicating header content in two or more CMSSW libraries. 


The requirement for module definitions of transitive includes is not new but we would like to reaffirm it is essential for modularization as it argues for the bottom-up approach. This approach can be challenging when there are a lot of dictionary header dependencies on external libraries. The dependencies under direct deployment control are easier to modularize as they require a single \emph{module.modulemap} file defining the module to be present at the base include location. Third-party dependencies that are not under deployment control are handled by ROOT's dictionary generator tool \emph{rootcling}. \emph{Rootcling} automatically creates a virtual file system overlay file and mounts the relevant modulemap file. The current implementation requires modification of the \emph{Cling} codebase but more configurable approaches are being investigated.

The current state of modularization of CMSSW is that over 70\% (190) of CMSSW libraries have corresponding C++ module files. Most of the core framework, \emph{FWCore}, data formats packages, \emph{DataForamts}, and condition/calibration classes, \emph{CondFormats}, packages have been modularized. The C++ modules reference (possibly transitively) external packages such as boost, tinyxl2, CLHEP and Eigen which are also being modularized.

\subsection{Performance Results}
\label{performance}

\begin{figure}[H]
\centering
\begin{subfigure}[t]{.49\linewidth}
\shadowimage[width=.9\textwidth]{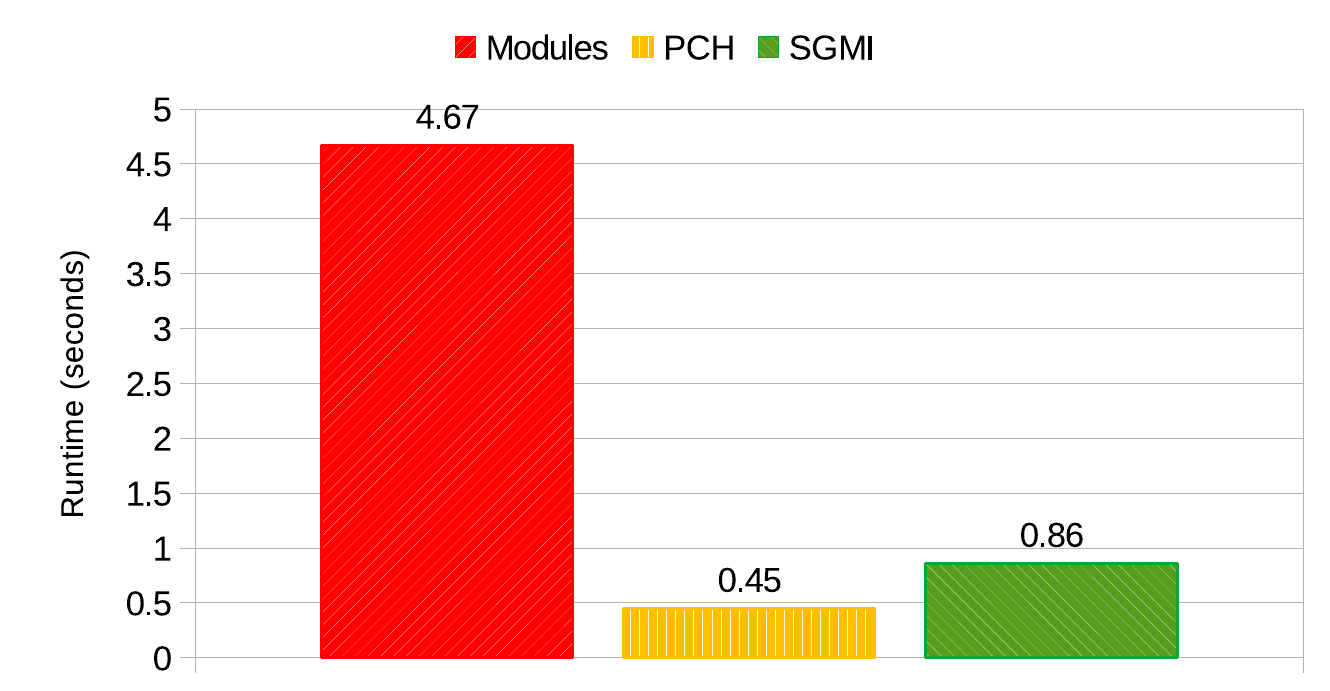}
\caption{Runtime} \label{fig:perf:3a}
\end{subfigure}
\begin{subfigure}[t]{.49\linewidth}
\shadowimage[width=.9\textwidth]{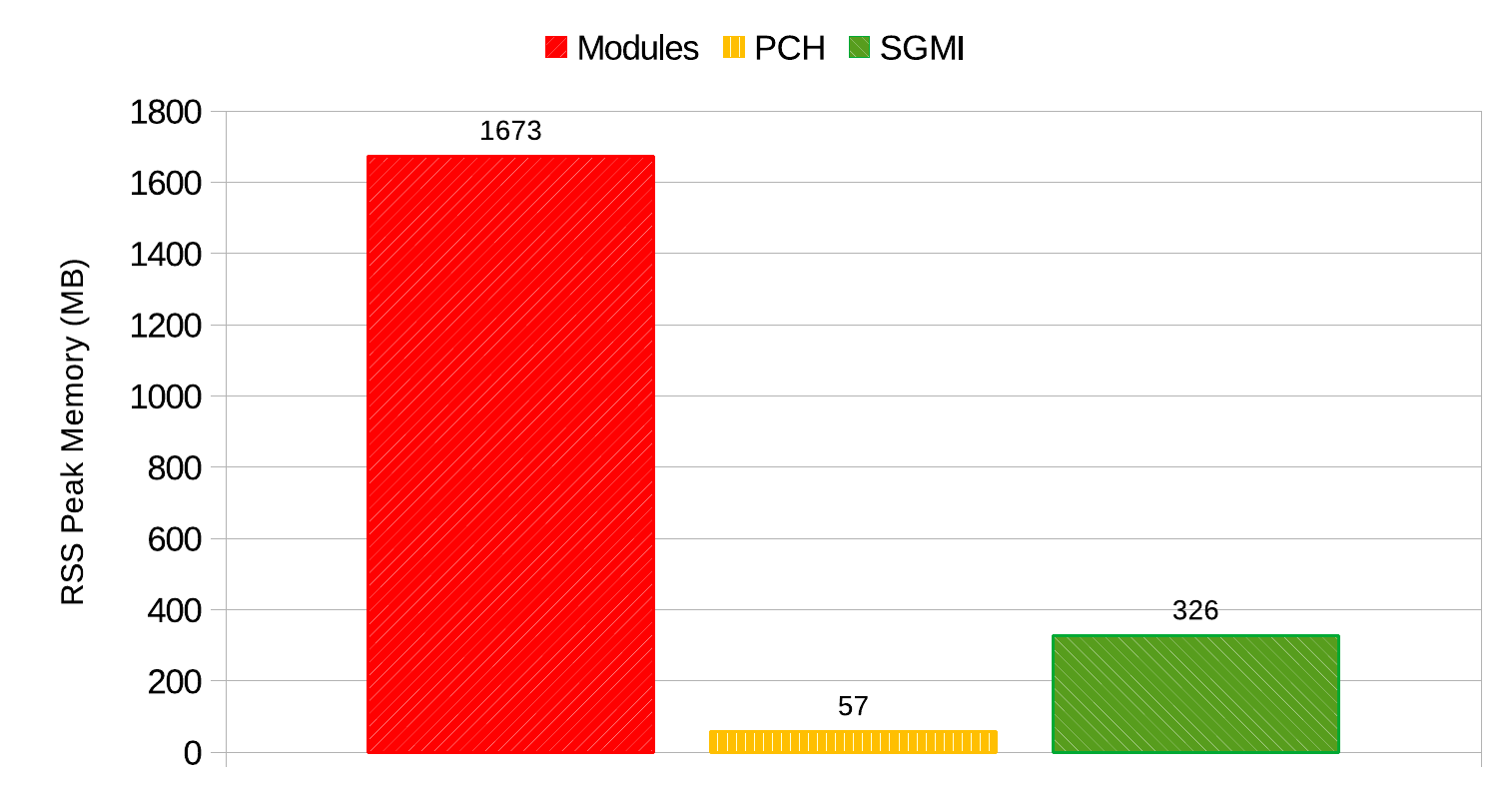}
\caption{Memory footprint} \label{fig:perf:3b}
\end{subfigure}
\caption{Performance benchmarks of \emph{root.exe -l -b -q} in CMSSW}\label{fig:gmiperfcmssw}
\end{figure}

At the time of writing of this paper we can only observe the behavior of the global module index in CMSSW indirectly by measuring the performance of the embedded in the ROOT distributed with CMSSW. This ROOT has access to 319 modules, around 190 are specific for the cmssw framework. The benchmark consists of running \emph{root.exe -l -b -q} in a configuration to preload all modules; using a PCH; and using an SGMI. The plots in Figure~\ref{fig:gmiperfcmssw} are done on CentOS 7.8.2003, kernel 3.10.0-1062.1.1.el7.x8\_64, Intel Broadwell, 30 GB RAM, 1 TB disk storage (CERN Virtual Machine Infrastructure).

Figure~\ref{fig:gmiperfcmssw} demonstrates the significant overhead in both memory footprint and runtime when all 319 modules are eagerly loaded. The impact of the overhead is unexpected because in a similar configuration standalone ROOT has only a 50 MB overhead in comparison to a PCH. This is due to the very well optimized modules which almost have no content duplication. Instead, the PCMs in cmssw currently contain many content duplication coming from the boost infrastructure. CMSSW has taken steps to reduce its dependence on boost in favor of using the available analogs in the new C++ standards. Despite the fact that for bigger workflows this startup overhead dissolves, it does not seem acceptable. On the other hand, the PCH shows very good results but as already mentioned it does not scale beyond ROOT.

The SGMI is not yet well optimized towards CMSSW. However we already see that it can improve performance for both memory and runtime relative to modules preloading. It is still suboptimal as it still eagerly loads 71 modules in advance without a real need. Reducing the eager loads in SGMI is a priority work item and will bring the performance close to the levels of the PCH for CMSSW the same way as we demonstrated for standalone ROOT in figure~\ref{fig:gmiperf}.

\section{Conclusion} \label{conclusion}

C++ modules are prominent technology which improves compilation speed by reducing content reparsing. The clang implementation used by ROOT is constantly improved by the LLVM community and will enable future performance benefits to the high-energy physics community.

Loading C++ module files currently has some limitations that introduce a non-negligible overhead which does not scale well in CMSSW. The paper outlines two strategies to overcome it -- using a global module index or frequently updating the Clang infrastructure to benefit from the upstream optimizations. The demonstrated performance benefits of SGMI are already implemented in ROOT and will be tested in the context of CMSSW to validate work in that direction.

CMSSW swiftly works towards adapting to newly introduced features in the C++ modules infrastructure in ROOT. Its modularization is ongoing and the necessary software stack is becoming more stable. After the modularization is complete we can more completely evaluate and improve the performance results.

\section{Acknowledgments}

The authors would like to thank Arpitha Raghunandan for her contributions to the global module index implementation and performance measurements as part of her work in the Google Summer of Code program 2019.

This work has been supported by an Intel Parallel Computing Center grant, by U.S. National Science Foundation grants OAC-1450377, OAC-1450323 and PHY-1624356, and by the U.S. Department of Energy, Office of Science.


\bibliography{main}

\end{document}